\documentclass[aps,prl,twocolumn,showpacs,preprintnumbers,amsmath,amssymb]{revtex4-1}
\usepackage{graphicx}
\usepackage[usenames]{color}
\begin{document}
%\preprint{NSF-KITP-12-027}

\title{Accurate exponents from approximate tensor renormalizations}
\author{Y. Meurice}
%\email[]{yannick-meurice@uiowa.edu}
\affiliation{Department of Physics and Astronomy\\ The University of Iowa\\
Iowa City, Iowa 52242, USA }

\def\lt{\lambda ^t}
\def\note{note}

\date{\today}
\begin{abstract}
We explain the recent numerical successes obtained by Tao Xiang's group, who developed and applied Tensor Renormalization Group methods for the Ising model on square and cubic lattices,  by the fact that their new truncation method 
sharply singles out a surprisingly small subspace of dimension two. We show that in the two-state approximation, their transformation can be handled analytically yielding a value 0.964  for the critical exponent $\nu$ much closer to the exact value 1 than 1.338 obtained in the Migdal-Kadanoff approximation.  We propose two alternative blocking procedures that preserve the isotropy and improve the accuracy to $\nu=0.987$ and 0.993 respectively.  We discuss applications to other 
classical lattice models, including models with fermions, and suggest that it could become a competitor for Monte Carlo methods suitable to calculate accurately critical exponents, take continuum limits and study near-conformal systems in arbitrarily large volumes.

\end{abstract}
\maketitle

%\section{Introduction}
The Renormalization Group (RG) ideas have triggered considerable conceptual and numerical progress in many branches of physics \cite{cardy96,Meurice:2011wy}. 
However, the basic method to thin down the number of degrees of freedom in configuration space \cite{wilson74}, often called ``block spinning'', has remained 
a formidable computational challenge for most classical lattice models (e. g., $O(N)$ spin models and  lattice gauge theories). 
A few years ago, inspired by the so-called tensor network states \cite{vc,cv} introduced in the context of the density matrix RG method \cite{PhysRevLett.69.2863,uli}, a Tensor RG (TRG) approach of two-dimensional (2D) classical lattice models was proposed \cite{PhysRevLett.99.120601}.   Successful 
approximations \cite{PhysRevLett.99.120601,PhysRevLett.103.160601,PhysRevB.81.174411} were found for the Ising model on honeycomb and triangular lattices. 

Very recently, the TRG method was successfully extended to the Ising model on square and cubic lattices by Tao Xiang's group \cite{PhysRevB.86.045139}. 
There are two important ingredients in their calculations. 
First, their formulation allows an {\it exact} block spinning procedure which 
separates neatly the degrees of freedom inside the block, which are integrated over, from those kept to communicate with the neighboring blocks. 
As explained below, this provides a more systematic way to implement ideas initiated by Migdal \cite{Migdal:1975zf} and 
Kadanoff \cite{Kadanoff:1976jb} (abbreviated as MK hereafter). The indices of the tensors run over some finite vector space of ``states" associated with finite volume link configurations. 
Second, they used a new method, based on higher order singular value decomposition, which selects in a very economical way the most important states that insure the communication among the blocks. Calculations using of the order of 20 states can be carried on a laptop computer. The excellent agreement  found with the Onsager solution in 2D 
for arbitrarily large volume suggests that TRG-based methods could become competitors for conventional Monte Carlo methods. 

In this Letter,   we show that the truncation method of Ref. \cite{PhysRevB.86.045139} for the 2D Ising model sharply singles out  a two-dimensional subspace of states. Keeping only these two states, we show that we can construct approximate RG transformations with 3 or 4 parameters, find the nontrivial fixed point and 
obtain precise estimates of the critical exponent $\nu$ associated with the correlation length from a linear analysis. The accuracy of the estimates is significantly better than for textbook examples such as the MK approximation \cite{Migdal:1975zf,Kadanoff:1976jb,Martinelli:1980sr}, the so-called approximate recursion formula \cite{wilson71b}  or other hierarchical approximations
\cite{baker72,hmreview}. 

The TRG formulation for the Ising model can be extended to $O(N)$ nonlinear sigma models and recent numerical implementations for $O(2)$ \cite{xiangprogress} indicate an optimistic outlook. It seems also possible to formulate models with local invariance and avoid sign problems. 
In this context, it is 
important to understand why the method works so well for the Ising model.

 The paper is organized as follows. We review the basic TRG formulation for  the Ising model on a square lattice emphasizing the connection with the MK ideas.  We then consider the cases of an isotropic blocking (as in the Migdal recursion) and an anisotropic blocking (as in the Kadanoff version and in Ref. \cite{PhysRevB.86.045139}).  
We also propose a new type of accurate isotropic projection based 
on a transfer matrix. We briefly discuss the 3D Ising model and how the method can be applied for lattice fermions.  The implications for the study of near-conformal systems and the calculations of critical exponents are discussed in the conclusions. 

We consider the nearest neighbor Ising model on a square lattice with an inverse temperature $\beta$. For easy reference, we stay close to the notations of Ref. \cite{PhysRevB.86.045139} where it is shown that the partition function can be written as 
\begin{equation}
Z=Tr\prod_{i}T^{(i)}_{xx'yy'} \ .
\label{eq:Z}
\end{equation}
The tensor $T^{(i)}_{xx'yy'}$ is attached to each site $i$ and Tr is a short notation for contractions over the links joining nearest neighbors on the lattice. 
The horizontal indices $x,\ x'$ and vertical indices $y,\ y'$ take the values 0 and 1. The tensor is zero for an odd number of 1. For an even number of 1, a factor $t^{1/2}$ 
(with $ t \equiv \tanh(\beta))$ appears for each 1 irrespectively of the direction. The partition function can be interpreted as a sum over intermediate states attached to the links. The reader familiar with the high temperature expansion of the model will recognize that this expression reproduces exactly the proper closed paths with the proper weights.  

\begin{figure}[h]
\includegraphics[width=2.3in,angle=0]{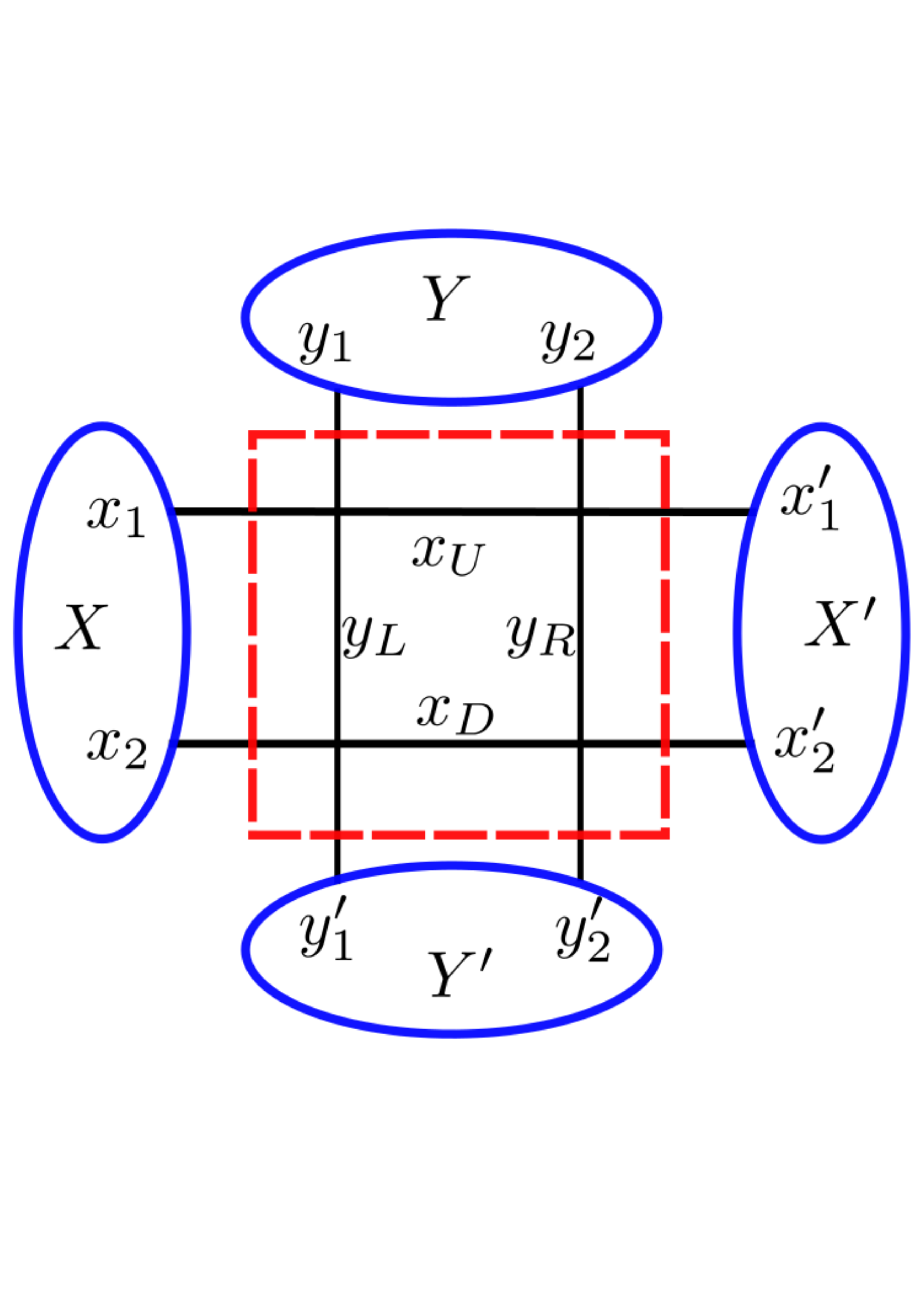}
\vskip-50pt
\includegraphics[width=2.3in,angle=0]{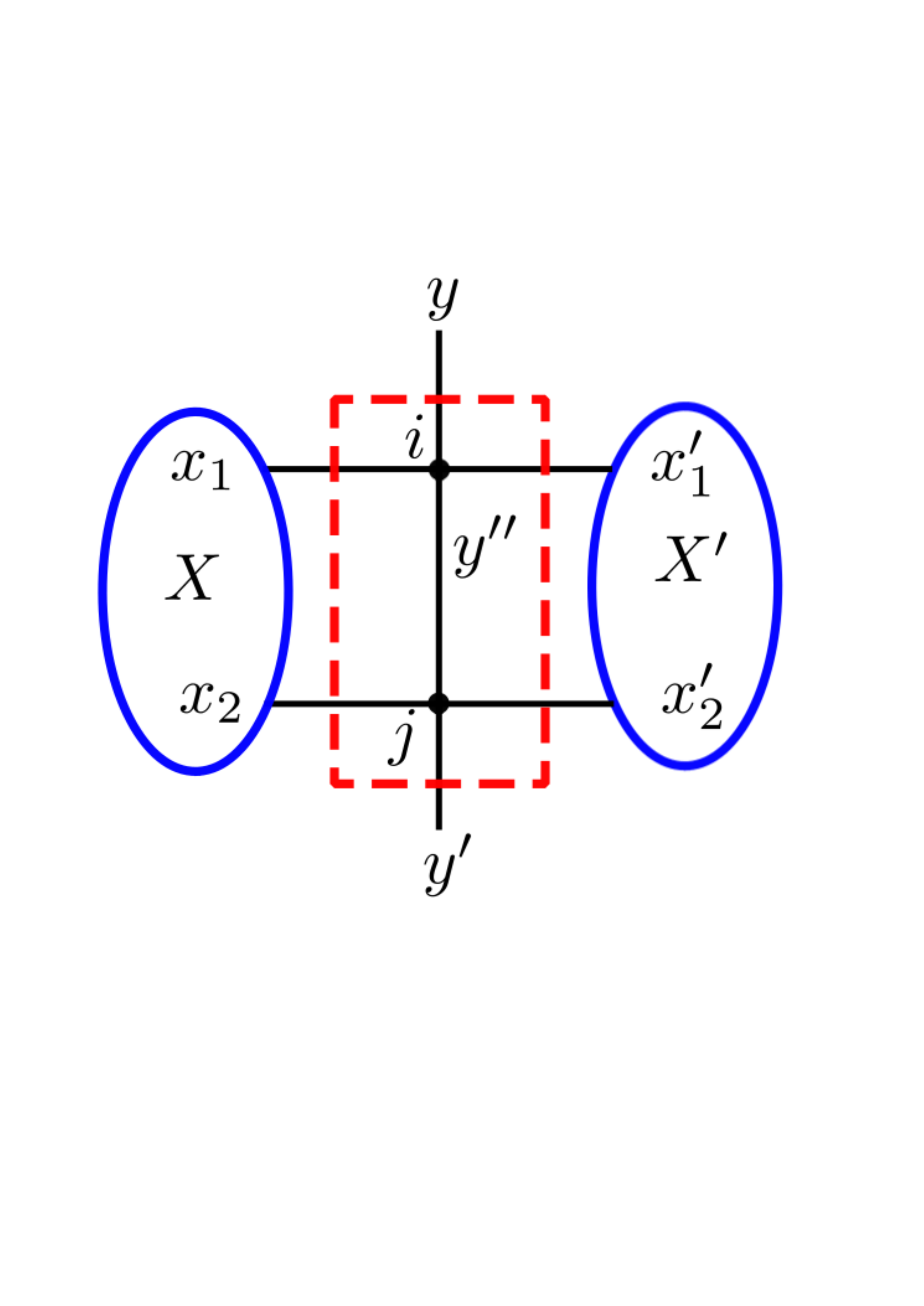}
\vskip-80pt
\caption{Graphical representation of $T'_{XX'YY'}$ (top) and $M ^{<ij>}_{X(x_1,x_2)X'(x_1'x_2')yy'}$ (bottom), the boundary of the block is represented by the dash lines and the product states by ovals. 
\label{fig:square} }
\end{figure}

We now use this reformulation to blockspin. 
First, we follow Migdal \cite{Migdal:1975zf} by using an isotropic procedure. We consider a square block 
enclosing 4 sites and sum over the states, inside the block, associated with the nearest neighbor links joining these 4 points. This defines a 
new rank 4 tensor $T'_{XX'YY'}$ where each index now takes 4 values. 
\begin{eqnarray}
&\ &T'_{X({x_1},{x_2})X'(x_1',x_2')Y(y_1,y_2)Y'(y_1',y_2')} = \\ \nonumber
&\ &\sum_{x_U,x_D,x_R,x_L}T_{x_1 x_U yy_L}T_{x_Ux_1'y_2y_R}T_{x_Dx_2'y_R y_2'}T_{x_2x_Dy_Ly_1'}\  ,
\end{eqnarray}
where $X(x_2,x_2)$ is a notation for the product states. For reasons that will become clear later, we use the convention: 
$X(0,0)=1,\  X(1,1)=1, \  X(1,0)=3,\  X(0,1)=4$. Later, we also use the ket notation $|00\rangle$ for $X=1$ etc..
This is represented graphically 
in Fig.  \ref{fig:square}. 
The new tensor can be used to define an {\it exact} expression of the partition function of the same form as (\ref{eq:Z}), however 
the number of states proliferates as $2^{2^n}$ after $n$ steps and approximations are needed in order to get an expression useful for practical purposes. 

The truncation method of Ref. \cite{PhysRevB.86.045139} relies on an 
anisotropic blocking involving two sites as shown at the bottom of Fig. \ref{fig:square}. 
This provides a new rank-4 tensor:
\begin{equation}
\label{eq:vblock}
M ^{<ij>}_{X(x_1,x_2)X'(x_1'x_2')yy'}=\sum_{y''}T^{(i)}_{x_1,x_1',y,y''}T^{(j)}_{x_2x_2'y''y'} \ ,
\end{equation}
which can be put in a canonical form by using a Higher Order Singular Value Decomposition defined by a unitary transformation on each of the four indices (see Ref. \cite{PhysRevB.86.045139} for justifications and refinements). The unitary transformation for each index is the one that diagonalizes the symmetric tensor obtained 
by summing over all the other indices in the following way: 
\begin{equation}
\label{eq:metric}
G_{XX'}=\sum_{X''yy'}M_{XX''yy'}M^*_{X'X''yy'} \ . 
\end{equation}

\begin{figure}[h]
\hskip-30pt
\includegraphics[width=2.8in,angle=0]{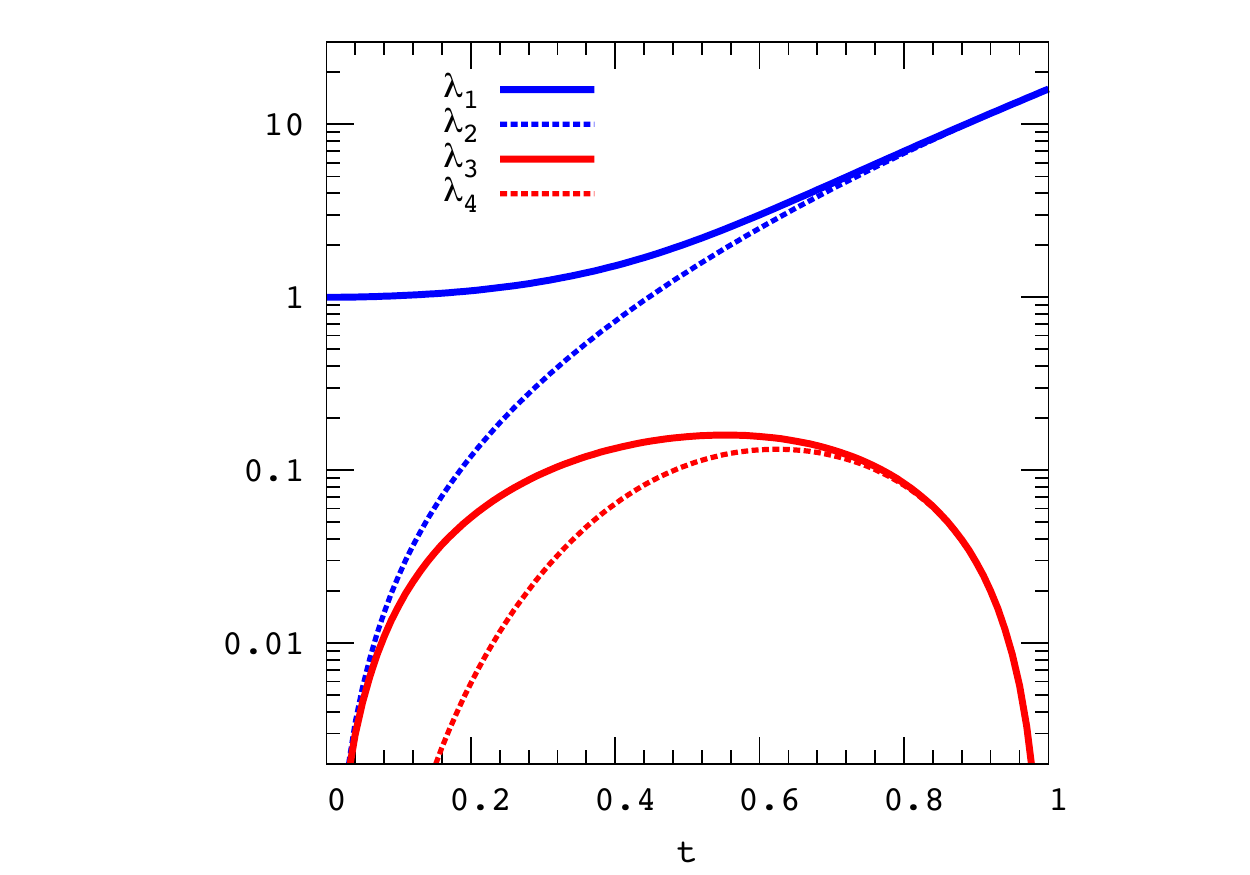}
\caption{The four eigenvalues of $G_{XX'}$ at the first step, on a logarithmic scale, as a function of $t=th\beta$.
\label{fig:logeig} }
\end{figure}

We now consider approximations where for each index we only keep the 
two states which correspond to the two largest eigenvalues of $G_{XX'}$. With our convention on the product states, this matrix is block diagonal because it does not connect states with even number of 1's  ($X=1,2$) to states with odd number of 1's ($X=3,4$). Numerically, we found that the two smallest eigenvalues are always very small compared to the largest one for any value of $\beta$ and that the second largest one is small at small $\beta$ and almost as large as the largest one at large $\beta$. This situation is illustrated for the initial step in Fig. \ref{fig:logeig}. 
After iterations the gap sharpens as if going to smaller $t$ for $t<t_c$ or larger $t$ for $t>t_c$.
The two new states have the form
\begin{eqnarray}
\label{eq:2states}
|0'\rangle &=&\cos\phi|00\rangle + \sin\phi |11\rangle \\ \nonumber
|1'\rangle &=& (|10\rangle + |01\rangle )/ \sqrt{2} \ .
\end{eqnarray} 
The angle $\phi$ is obtained by diagonalizing the even-even block. The symmetric form of $|1'\rangle$ is a consequence of 
$G_{33}=G_{44}$, itself due to reflection symmetry. 

With this rather drastic projection, we obtain a new rank 4 tensor with indices taking again two values and the same parity selection rule. We divide the new tensor by a constant in such a way that, dropping all the primes herafter, $T_{0000}=1$ . In the isotropic case, the reflection symmetry imposes that $T_{1010}=T_{0110}=T_{1001}=T_{0101}\equiv t_1$, $T_{1100}=T_{0011}\equiv t_2$, $T_{1111}\equiv t_3$. For the initial tensor, we have what we call later the ``Ising condition":\begin{equation}
t_1=t_2\  {\rm and} \  t_3=t_1^2\ ,
\label{eq:ic}
\end{equation}but this property is not preserved by the blocking procedure which can be expressed as a mapping of the three dimensional parameter space $(t_1,t_2,t_3)$ into itself. This can be expressed as rational expressions involving the values of the tensor and trigonometric functions of $\phi$. They are analytical expressions and derivatives can be taken to calculate the linearized map. They are not written explicitly here but can be obtained easily with symbolic manipulation programs. 

%The explicit form can be worked out easily by hand or using symbolic manipulation. 
A fixed point is found for $\beta=0.411594547\dots$ not far from the exact value 0.44068$\dots$.
The fixed point is approximately $t_1^\star=0.443492,t_2^\star = 0.298462,t_3^\star=0.304291$ and shows significant departure from the Ising condition (\ref{eq:ic}). 
The eigenvalues of the linearized RG transformation are 
$\lambda_1$=2.0177, $\lambda_2$=0.2022 and $\lambda_3$=0.09967. The scaling factor $b$ is 2, and this implies the values for the critical exponents $\nu =\log b/\log \lambda_1\simeq 0.987$ and $\omega=-\log \lambda_2/\log b \simeq 2.31$ which can be compared with the exact values 1 and 2 respectively. 

For comparison, the also isotropic Migdal recursion for the $D=2$ Ising model and $b=2$ can be written as 
\begin{equation}
t'=2t^2/(1+t^4)
\label{eq:migdal}
\end{equation}
The fixed point is at $t^\star=0.543689$ and  corresponds to $\beta_c=0.609378$ and $\nu$ = 1.338. This approximation can be seen as a two-state approximation which in addition 
requires the Ising condition. Improvements of MK discussed in Ref.  \cite{Martinelli:1980sr} lead to a value of $\nu$  = 0.796 at first order in their expansion parameter and 0.93(1) at second order, the error bar coming from the use of different Pad\'e approximants.

This suggests that the Ising condition is too restrictive. It is nevertheless possible to modify the angle $\theta$ in Eq. (\ref{eq:2states}) in such a way that Ising condition remains valid. This is nontrivial because the Ising condition amounts to two equations. However, an explicit calculation shows that if 
$\tan \phi = 1/\tan \beta$ the {\it two} conditions are satisfied and the mapping takes the form
\begin{equation}
t'=(1+t^2)/2t \ ,
\end{equation}
which unfortunately has only the low temperature fixed point. 

The above results can be compared with the two-state truncation for the anisotropic blocking used in  Ref. \cite{PhysRevB.86.045139} where a first projection occurs after blocking pairs of vertical sites as already introduced in Eq. (\ref{eq:vblock}). The complete 
transformation is then obtained by repeating the procedure with a horizontal blocking as in Kadanoff's proposal  \cite{Kadanoff:1976jb}. 
We use our previous notations but with $h$ and $v$ subscripts denoting the horizontal and vertical couplings respectively: $T_{1100}\equiv t_{2h}$ and $T_{0011}\equiv t_{2v}$. We keep the same notations for $t_1$ and $t_3$ which are invariant under the interchange of the vertical and horizontal directions. We have now a map with four parameters. In the Ising condition, we need to replace $t_2$ by $\sqrt{t_{2v}t_{2h}}$. 
We take initial values that satisfy the Ising condition and are isotropic ($t_{2h}=t_{2v}=t=\tanh(\beta)$). The critical value is  then $\beta_c =0.37945324441109\dots$ 
with a nontrivial fixed point for approximately $t_{2h}^\star$=0.21358, $t_{2v}^\star$=0.37924, $t_1^\star$=0.41998 and $t_3^\star$= 0.27177, which is clearly anisotropic and violates the Ising condition. We have an additional, intermediate, eigenvalue which is approximately 0.657 and specific to the anisotropic case. This value is not very far from unity which is why it requires more fine-tuning of $\beta$ to get rid of the irrelevant directions. The first and third eigenvalues are 2.052 and 0.1934 respectively which implies $\nu$= 0.964 and $\omega$= 2.37.

This anisotropic version can be compared with the Kadanoff recursion \cite{Kadanoff:1976jb} 
for $b$ =2, where 
first, the horizontal bonds are slided vertically with $\beta_h$ doubled and $t_v$ squared. 
This corresponds to doubling the vertical lattice spacing first as we did above. 
After repeating the procedure with horizontal moves, we
obtain the Migdal recursion of Eq. (\ref{eq:migdal}) for $t_v$ while for $t_h$ we obtain 
\begin{equation}
t'_h=(2t_h/(1+t_h^2))^2\ ,
\label{eq:kad}
\end{equation}
which corresponds to reversing the order of the two operations. 
The fixed point  is  $t_v^\star=0.543689$  and $t_h^\star=(t_v^\star)^2$=0.295598. The eigenvalue is the same in both directions and the value of $\nu$ 
identical to the Migdal case. 

We have also considered the isotropic map with a different truncation. Instead of Eq. (\ref{eq:metric}), we use
\begin{equation}
\label{eq:metric2}
\tilde{G}_{XX'}=\sum_{X''yy'}M_{XX''yy}M^*_{X'X''y'y'} \ . 
\end{equation}
The trace of this matrix is the partition function for a $2 \times 2$ model with periodic boundary conditions. 
This gives a slightly displaced fixed point at $\beta_c =0.3948678$. 
$t_1^\star$=0.42229, $t_2^\star$=0.28637 and $t_3^\star$=0.27466. The values of the exponents are $\nu=0.993$ and $\omega=2.37$. 
It should be noted that in this case the two small eigenvalues at criticality (0.00128 and 0.0000698) are much smaller than in the first calculation (0.118 and 0.0525) which may explain the improved accuracy on $\nu$. 

Extensions of these methods for more states, more components and more dimensions are in progress. 
For practical purposes, the analytical methods discussed above need to be implemented numerically.
We have succeeded to reproduce all the results obtained so far with adequate accuracy using numerical procedures which can be 
implemented using the most common programming languages. The fixed point can be found by monitoring successive bifurcations 
in tensor values. The high and low temperature phases are characterized by the fact that some  tensor values go to zero (when $\beta$ is too small) or one (when $\beta$ is too large) if we iterate enough times.  We can then fine-tune $\beta$ and observe the stabilization of tensors at some nontrivial  values. The eigenvalues can be found by taking numerical derivatives of the one step transformation 
with respect to the initial values as close as possible to the nontrivial fixed point. This requires variations small enough but not too small
since  we have only limited accuracy on $\beta_c$, however 
accurate results can be obtained using linear extrapolations to zero variation.

The formulation can be extended in $D$ dimension using tensors with with $2D$ indices. 
The reason the TRG blocking works well in any dimension is that the links are orthogonal (dual) to domain boundaries. 
We keep the link variables across the boundaries of the block fixed and sum unrestrictedly over all the states inside the block.
We have extended the third method described above to the 3D Ising model. The block is then a cube with four external legs coming out of each of  the 
six faces. The blocking and the projection can be built out of the loop made by the four edges of a face with 4 external legs attached to each of the 4 corners. 
The initial transfer matrix can be obtained by tracing the external legs in the plane of the loop with their opposite leg. We then obtain a $16\times16$ matrix  corresponding to the 4 legs coming 
out of the plane of the loop in each direction. This matrix splits  into two $8\times8$ blocks. Numerically, we found $\beta_c=0.1996597773239\dots$ for a two-state projection. The initial eigenvalues of 
$\tilde{G}_{XX'}$ at $\beta_c$ are 1.2325, 0.5082 and a pair with value 0.1682 which cannot be considered as small. For this reason, the value of  $\nu \simeq 0.74$ for the 2D Ising model in the two-states approximation is not very close to the accurate value 0.630(2) \cite{pelissetto00b} but nevertheless more accurate than the MK approximation value 1.055.  
%A 4-state calculation cannot be done analytically and a numerical implementation is in progress \cite{judahprogress}. 

There is good empirical evidence  \cite{PhysRevB.86.045139,xiangprogress} supporting the idea that as we increase the number of states in TRG calculations, the numerical estimates of energy and entropy 
get closer to values obtained by exact methods (for the 2D Ising model) or Monte Carlo simulations (for the 2D $O(2)$ model). However, in order to reach a reasonably large number of states (20-30) on a laptop, 
one needs to use efficient methods. The anisotropic TRG methods discussed above  involve less contractions or external legs and the computational cost for $N_s$ states can be limited \cite{PhysRevB.86.045139} to 
a $N_s^7$ growth in two dimensions and a $N_s^{11}$ growth in three dimensions.  Numerical implementations of our method for up to 12 states are now in progress \cite{judahprogress}. In all cases, we observe a sharp split between a few large eigenvalues, their number being characteristic of the phase, and the small ones as in the two states case. The interesting behavior of these new maps will be discussed elsewhere \cite{judahprogress}.

Because of the binary nature of Grassmann numbers, the techniques developed for Ising models can also be used for lattice models with fermions. In the case where we have 
Grassmann variables $\psi^{(i)}_\alpha$ with $\alpha =1, 2, \dots,q$ at every site $i$ and nearest neighbor interactions specified by a $q\times q$ matrix $A_{\alpha\beta}^{<ij>}$ at every link $<ij>$, we can write
\begin{equation}
\exp (\psi^{(i)}_\alpha A_{\alpha\beta}^{<ij>}\psi ^{(j)}_{\beta})=
\sum_{n^{<ij>}_{\alpha}=0,1}\prod_{\alpha=1}^q (\tilde\psi ^{(i)}_\alpha \lambda_\alpha\tilde\psi ^{(j)}_\alpha )^{n^{<ij>}_{\alpha}}\ . \nonumber
\end{equation}
The $\tilde\psi$ are linear combinations of the $\psi$ at the same site obtained from the decomposition $A=U\Lambda V^{\dagger}$ with $U$ and $V$ unitary and $\Lambda$ diagonal with elements ${\lambda_{\alpha}}$. The terms can then be factorized at every site and the local integrations performed. 
The states are now parametrized by ${n^{<ij>}_{\alpha}}$ and there are $2^q$ of them at every link. Translational invariance is essential to perform large volume calculations. 
For this reason, possible gauge interations would need to be averaged inside the blocks. 

In summary, we have shown that two-state approximations of the TRG capture the universal behavior of Ising models much better than the MK approximation. 
Building on the numerical success of Ref. \cite{PhysRevB.86.045139} which uses more states, we expect to be able to use the methods presented here to calculate the 
exponents of the 3D Ising model with unprecedented accuracy and study the analytical picture of the critical behavior provided in Ref. \cite{PhysRevD.86.025022,ElShowk:2012hu}. 
Recent numerical results for the TRG method applied to the $O(2)$ model \cite{xiangprogress} suggest that improvements of the MK approximation  could be applied to abelian models and resolve the controversy regarding the confinement in 4D $U(1)$ gauge theory discussed in Ref. \cite{Tomboulis:2009zz}. We are hoping to be able to extend the TRG method for lattice gauge theories with fermions. Being able to block spin accurately would provide an efficient tool to study the continuum limits of asymptotically free models and the conformal window of models that could provide alternatives to the fundamental Higgs mechanism \cite{DeGrand:2010ba}.

Acknowledgments. Our work on the subject started  while attending the KITPC workshop ``Critical Properties of Lattice Models" in summer 2012. 
We thank  M.C. Banuls, S. Chandrasekharan, A. Denbleyker, A. Li, Y. Liu, Y, M. Ogilvie, P. Orland, M. Qin, T. Tomboulis, J. Unmuth-Yockey, X-G Wen, T. Xiang, Z. Xie, J. Yu, and H. Zou for valuable conversations and suggestions. 
This research was supported in part  by the Department of Energy
under Contract No. FG02-91ER40664  %\end{acknowledgments}


\begin{thebibliography}{10}%
\makeatletter
\providecommand \@ifxundefined [1]{%
 \ifx #1\undefined \expandafter \@firstoftwo
 \else \expandafter \@secondoftwo
\fi
}%
\providecommand \@ifnum [1]{%
 \ifnum #1\expandafter \@firstoftwo
 \else \expandafter \@secondoftwo
\fi
}%
\providecommand \enquote [1]{``#1''}%
\providecommand \bibnamefont  [1]{#1}%
\providecommand \bibfnamefont [1]{#1}%
\providecommand \citenamefont [1]{#1}%
\providecommand\href[0]{\@sanitize\@href}%
\providecommand\@href[1]{\endgroup\@@startlink{#1}\endgroup\@@href}%
\providecommand\@@href[1]{#1\@@endlink}%
\providecommand \@sanitize [0]{\begingroup\catcode`\&12\catcode`\#12\relax}%
\@ifxundefined \pdfoutput {\@firstoftwo}{%
 \@ifnum{\z@=\pdfoutput}{\@firstoftwo}{\@secondoftwo}%
}{%
 \providecommand\@@startlink[1]{\leavevmode}%
 \providecommand\@@endlink[0]{}%
}{%
 \providecommand\@@startlink[1]{%
  \leavevmode
  \pdfstartlink
   attr{/Border[0 0 1 ]/H/I/C[0 1 1]}%
   user{/Subtype/Link/A<</Type/Action/S/URI/URI(#1)>>}%
  \relax
 }%
 \providecommand\@@endlink[0]{\pdfendlink}%
}%
\providecommand \url  [0]{\begingroup\@sanitize \@url }%
\providecommand \@url [1]{\endgroup\@href {#1}{\urlprefix}}%
\providecommand \urlprefix [0]{URL }%
\providecommand \Eprint[0]{\href }%
\@ifxundefined \urlstyle {%
  \providecommand \doi [1]{doi:\discretionary{}{}{}#1}%
}{%
  \providecommand \doi [0]{doi:\discretionary{}{}{}\begingroup
  \urlstyle{rm}\Url }%
}%
\providecommand \doibase [0]{http://dx.doi.org/}%
\providecommand \Doi[1]{\href{\doibase#1}}%
\providecommand \bibAnnote [3]{%
  \BibitemShut{#1}%
  \begin{quotation}\noindent
    \textsc{Key:}\ #2\\\textsc{Annotation:}\ #3%
  \end{quotation}%
}%
\providecommand \bibAnnoteFile [2]{%
  \IfFileExists{#2}{\bibAnnote {#1} {#2} {\input{#2}}}{}%
}%
\providecommand \typeout [0]{\immediate \write \m@ne }%
\providecommand \selectlanguage [0]{\@gobble}%
\providecommand \bibinfo [0]{\@secondoftwo}%
\providecommand \bibfield [0]{\@secondoftwo}%
\providecommand \translation [1]{[#1]}%
\providecommand \BibitemOpen[0]{}%
\providecommand \bibitemStop [0]{}%
\providecommand \bibitemNoStop [0]{.\EOS\space}%
\providecommand \EOS [0]{\spacefactor3000\relax}%
\providecommand \BibitemShut [1]{\csname bibitem#1\endcsname}%
%</preamble>
\bibitem{cardy96}%
  \BibitemOpen
  \bibfield{author}{%
  \bibinfo {author} {\bibfnamefont{J.~L.}\ \bibnamefont{Cardy}}\ }%
  \bibinfo {note} {{\it Scaling and Renormalization in Statistical Physics}, (Cambridge Univ. Pr., Cambridge, 1996)}%
  \bibAnnoteFile{NoStop}{cardy96}%
  \bibitem{Meurice:2011wy}%
  \BibitemOpen
  \bibfield{author}{%
  \bibinfo {author} {\bibfnamefont{Y.}~\bibnamefont{Meurice}}, \bibinfo
  {author} {\bibfnamefont{R.}~\bibnamefont{Perry}},\ and\ \bibinfo {author}
  {\bibfnamefont{S.-W.}\ \bibnamefont{Tsai}},\ }%
  \bibfield{journal}{%
  \Doi{10.1098/rsta.2011.0117}{\bibinfo {journal} {Phil. Trans. Roy. Soc.}}\
  }%
  \textbf{\bibinfo {volume} {A369}},\ \bibinfo {pages} {2602} (\bibinfo {year}
  {2011}) and Refs. therein%\ \Eprint{http://arxiv.org/abs/1102.5717}{arXiv:1102.5717 [hep-th]}%
  \bibAnnoteFile{NoStop}{Meurice:2011wy}%
%%CITATION = ARXIV:1102.5717;%%
\bibitem{wilson74}%
  \BibitemOpen
  \bibfield{author}{%
  \bibinfo {author} {\bibfnamefont{K.}~\bibnamefont{Wilson}}\ and\ \bibinfo
  {author} {\bibfnamefont{J.}~\bibnamefont{Kogut}},\ }%
  \bibfield{journal}{%
  \bibinfo {journal} {Phys.\ Rep.}\ }%
  \textbf{\bibinfo {volume} {12}},\ \bibinfo {pages} {75} (\bibinfo {year}
  {1974})%
  \bibAnnoteFile{NoStop}{wilson74}%
\bibitem{vc}%
  \BibitemOpen
  \bibfield{author}{%
  \bibinfo {author} {\bibfnamefont{F.}~\bibnamefont{{Verstraete}}}\ and\
  \bibinfo {author} {\bibfnamefont{J.~I.}\ \bibnamefont{{Cirac}}},\ }%
  \bibfield{journal}{%
  \bibinfo {journal} {cond-mat/0407066}\ }%
   (\bibinfo {year} {2004})%,\
 % \Eprint{http://arxiv.org/abs/arXiv:cond-mat/0407066}{arXiv:cond-mat/0407066}%
  \bibAnnoteFile{NoStop}{vc}%
\bibitem{cv}%
  \BibitemOpen
  \bibfield{author}{%
  \bibinfo {author} {\bibfnamefont{J.~I.}\ \bibnamefont{{Cirac}}}\ and\
  \bibinfo {author} {\bibfnamefont{F.}~\bibnamefont{{Verstraete}}},\ }%
  \bibfield{journal}{%
  \Doi{10.1088/1751-8113/42/50/504004}{\bibinfo {journal} {J. of Phys. A
%  Mathematical General
  }}\ }%
  \textbf{\bibinfo {volume} {42}},\ \bibinfo {pages} {4004} (%\bibinfo {month}
  %{Dec.}\ 
  \bibinfo {year} {2009})%,\
  %\Eprint{http://arxiv.org/abs/0910.1130}{arXiv:0910.1130 [cond-mat.str-el]}%
  \bibAnnoteFile{NoStop}{cv}%
  \bibitem{PhysRevLett.69.2863}%
  \BibitemOpen
  \bibfield{author}{%
  \bibinfo {author} {\bibfnamefont{S.~R.}\ \bibnamefont{White}},\ }%
  \bibfield{journal}{%
  \Doi{10.1103/PhysRevLett.69.2863}{\bibinfo {journal} {Phys. Rev. Lett.}}\ }%
  \textbf{\bibinfo {volume} {69}},\ \bibinfo {pages} {2863} (%\bibinfo {month}
  %{Nov}\ 
  \bibinfo {year} {1992})%,\
  %\url{http://link.aps.org/doi/10.1103/PhysRevLett.69.2863}%
  \bibAnnoteFile{NoStop}{PhysRevLett.69.2863}%
\bibitem{uli}%
  \BibitemOpen
  \bibfield{author}{%
  \bibinfo {author} {\bibfnamefont{U.}~\bibnamefont{Schollwoeck}},\ }%
  \bibfield{journal}{%
  \bibinfo {journal} {Phil. Trans. R. Soc. A}\ }%
  \textbf{\bibinfo {volume} {369}},\ \bibinfo {pages} {2643} (\bibinfo {year}
  {2011})%
  \bibAnnoteFile{NoStop}{uli}%
%%CITATION = 1010.4741;%%
\bibitem{PhysRevLett.99.120601}%
  \BibitemOpen
  \bibfield{author}{%
  \bibinfo {author} {\bibfnamefont{M.}~\bibnamefont{Levin}}\ and\ \bibinfo
  {author} {\bibfnamefont{C.~P.}\ \bibnamefont{Nave}},\ }%
  \bibfield{journal}{%
  \Doi{10.1103/PhysRevLett.99.120601}{\bibinfo {journal} {Phys. Rev. Lett.}}\
  }%
  \textbf{\bibinfo {volume} {99}},\ \bibinfo {pages} {120601} (%\bibinfo {month}
 % {Sep}\ 
  \bibinfo {year} {2007})%
  %\url{http://link.aps.org/doi/10.1103/PhysRevLett.99.120601}%
  \bibAnnoteFile{NoStop}{PhysRevLett.99.120601}%
  \bibitem{PhysRevLett.103.160601}%
  \BibitemOpen
  \bibfield{author}{%
  \bibinfo {author} {\bibfnamefont{Z.~Y.}\ \bibnamefont{Xie}}, \bibinfo
  {author} {\bibfnamefont{H.~C.}\ \bibnamefont{Jiang}}, \bibinfo {author}
  {\bibfnamefont{Q.~N.}\ \bibnamefont{Chen}}, \bibinfo {author}
  {\bibfnamefont{Z.~Y.}\ \bibnamefont{Weng}},\ and\ \bibinfo {author}
  {\bibfnamefont{T.}~\bibnamefont{Xiang}},\ }%
  \bibfield{journal}{%
  \Doi{10.1103/PhysRevLett.103.160601}{\bibinfo {journal} {Phys. Rev. Lett.}}\
  }%
  \textbf{\bibinfo {volume} {103}},\ \bibinfo {pages} {160601} (\bibinfo {year} {2009})%\
  %\url{http://link.aps.org/doi/10.1103/PhysRevLett.103.160601}%
  \bibAnnoteFile{NoStop}{PhysRevLett.103.160601}%
\bibitem{PhysRevB.81.174411}%
  \BibitemOpen
  \bibfield{author}{%
  \bibinfo {author} {\bibfnamefont{H.~H.}\ \bibnamefont{Zhao}}, \bibinfo
  {author} {\bibfnamefont{Z.~Y.}\ \bibnamefont{Xie}}, \bibinfo {author}
  {\bibfnamefont{Q.~N.}\ \bibnamefont{Chen}}, \bibinfo {author}
  {\bibfnamefont{Z.~C.}\ \bibnamefont{Wei}}, \bibinfo {author}
  {\bibfnamefont{J.~W.}\ \bibnamefont{Cai}},\ and\ \bibinfo {author}
  {\bibfnamefont{T.}~\bibnamefont{Xiang}},\ }%
  \bibfield{journal}{%
  \Doi{10.1103/PhysRevB.81.174411}{\bibinfo {journal} {Phys. Rev. B}}\ }%
  \textbf{\bibinfo {volume} {81}},\ \bibinfo {pages} {174411} (\bibinfo {year} {2010})%
  %\url{http://link.aps.org/doi/10.1103/PhysRevB.81.174411}%
  \bibAnnoteFile{NoStop}{PhysRevB.81.174411}%
  \bibitem{PhysRevB.86.045139}%
  \BibitemOpen
  \bibfield{author}{%
  \bibinfo {author} {\bibfnamefont{Z.~Y.}\ \bibnamefont{Xie}}, \bibinfo
  {author} {\bibfnamefont{J.}~\bibnamefont{Chen}}, \bibinfo {author}
  {\bibfnamefont{M.~P.}\ \bibnamefont{Qin}}, \bibinfo {author}
  {\bibfnamefont{J.~W.}\ \bibnamefont{Zhu}}, \bibinfo {author}
  {\bibfnamefont{L.~P.}\ \bibnamefont{Yang}},\ and\ \bibinfo {author}
  {\bibfnamefont{T.}~\bibnamefont{Xiang}},\ }%
  \bibfield{journal}{\Doi{10.1103/PhysRevB.86.045139}{\bibinfo {journal} {Phys. Rev. B}}\ }%
  \textbf{\bibinfo {volume} {86}},\ \bibinfo {pages} {045139} (%\bibinfo {month}
 % {Jul}\ 
  \bibinfo {year} {2012})
%  \url{http://link.aps.org/doi/10.1103/PhysRevB.86.045139}%
  \bibAnnoteFile{NoStop}{PhysRevB.86.045139}%
 \bibitem{Migdal:1975zf}%
  \BibitemOpen
  \bibfield{author}{%
  \bibinfo {author} {\bibfnamefont{A.~A.}\ \bibnamefont{Migdal}},\ }%
  \bibfield{journal}{%
  \bibinfo {journal} {Sov.Phys.JETP}\ }%
  \textbf{\bibinfo {volume} {42}},\ \bibinfo {pages} {743} (\bibinfo {year}
  {1975})%
  \bibAnnoteFile{NoStop}{Migdal:1975zf}%
%%CITATION = SPHJA,42,743;%%
\bibitem{Kadanoff:1976jb}%
  \BibitemOpen
  \bibfield{author}{%
  \bibinfo {author} {\bibfnamefont{L.}~\bibnamefont{Kadanoff}},\ }%
  \bibfield{journal}{%
  \Doi{10.1016/0003-4916(76)90066-X}{\bibinfo {journal} {Annals Phys.}}\ }%
  \textbf{\bibinfo {volume} {100}},\ \bibinfo {pages} {359} (\bibinfo {year}
  {1976})%
  \bibAnnoteFile{NoStop}{Kadanoff:1976jb}%
  %%CITATION = APNYA,100,359;%%
  \bibitem{Martinelli:1980sr}%
  \BibitemOpen
  \bibfield{author}{%
  \bibinfo {author} {\bibfnamefont{G.}~\bibnamefont{Martinelli}}\ and\ \bibinfo
  {author} {\bibfnamefont{G.}~\bibnamefont{Parisi}},\ }%
  \bibfield{journal}{%
  \Doi{10.1016/0550-3213(81)90415-6}{\bibinfo {journal} {Nucl.Phys.}}\ }%
  \textbf{\bibinfo {volume} {B180}},\ \bibinfo {pages} {201} (\bibinfo {year}
  {1981})%
  \bibAnnoteFile{NoStop}{Martinelli:1980sr}%
%%CITATION = NUPHA,B180,201;%%
\bibitem{wilson71b}%
  \BibitemOpen
  \bibfield{author}{%
  \bibinfo {author} {\bibfnamefont{K.}~\bibnamefont{Wilson}},\ }%
  \bibfield{journal}{%
  \bibinfo {journal} {Phys. Rev. B.}\ }%
  \textbf{\bibinfo {volume} {4}},\ \bibinfo {pages} {3184} (\bibinfo {year}
  {1971})%
  \bibAnnoteFile{NoStop}{wilson71b}%
\bibitem{baker72}%
  \BibitemOpen
  \bibfield{author}{%
  \bibinfo {author} {\bibfnamefont{G.}~\bibnamefont{Baker}},\ }%
  \bibfield{journal}{%
  \bibinfo {journal} {Phys.\ Rev.\ B}\ }%
  \textbf{\bibinfo {volume} {5}},\ \bibinfo {pages} {2622} (\bibinfo {year}
  {1972})%
  \bibAnnoteFile{NoStop}{baker72}%
\bibitem{hmreview}%
  \BibitemOpen For a review see: 
  \bibfield{author}{%
  \bibinfo {author} {\bibfnamefont{Y.}~\bibnamefont{Meurice}},\ }%
  \bibfield{journal}{%
  \bibinfo {journal} {J. Phys.}\ }%
  \textbf{\bibinfo {volume} {A40}},\ \bibinfo {pages} {R39} (\bibinfo {year}
  {2007})%\ \Eprint{http://arxiv.org/abs/hep-th/0701191}{hep-th/0701191}%
  \bibAnnoteFile{NoStop}{hmreview}%
%%CITATION = HEP-TH/0701191;%%
 \bibitem{xiangprogress}%
  \BibitemOpen
  \bibfield{author}{%
  \bibinfo {author} {\bibfnamefont{A.}~\bibnamefont{Denbleyker}}, \bibinfo
  {author} {\bibfnamefont{Y.}~\bibnamefont{Liu}}, \bibinfo {author}
  {\bibfnamefont{Y.}~\bibnamefont{Meurice}}, \bibinfo {author}
  {\bibfnamefont{M.}~\bibnamefont{Qin}}, \bibinfo {author}
  {\bibfnamefont{T.}~\bibnamefont{Xiang}}, \bibinfo {author}
  {\bibfnamefont{Z.}~\bibnamefont{Xie}},\ and\ \bibinfo {author}
  {\bibfnamefont{J.}~\bibnamefont{Yu}},\ }%
  \bibinfo {note} {preprint in progress}%\ %
  \bibAnnoteFile{NoStop}{xiangprogress}%
  \bibitem{pelissetto00b}%
  \BibitemOpen
  \bibfield{author}{%
  \bibinfo {author} {\bibfnamefont{A.}~\bibnamefont{Pelissetto}}\ and\ \bibinfo
  {author} {\bibfnamefont{E.}~\bibnamefont{Vicari}},\ }%
  \bibfield{journal}{%
  \bibinfo {journal} {Phys. Rept.}\ }%
  \textbf{\bibinfo {volume} {368}},\ \bibinfo {pages} {549} (\bibinfo {year}
  {2002})\ %\ \Eprint{http://arxiv.org/abs/cond-mat/0012164}
  {cond-mat/0012164}%
  \bibAnnoteFile{NoStop}{pelissetto00b}%
%%CITATION = COND-MAT 0012164;%%
\bibitem{judahprogress}%
  \BibitemOpen
  \bibfield{author}{%
  \bibinfo {author} {\bibfnamefont{Y.}~\bibnamefont{Meurice}},\  \bibinfo
  {author} {\bibfnamefont{J.}~\bibnamefont{Unmuth-Yockey}},\  and \ \bibinfo
  {author} {\bibfnamefont{H.}~\bibnamefont{Zou}}},\ %
  \bibinfo {title} {work in progress}%
  \bibAnnoteFile{NoStop}{judahprogress}%
    \bibitem{PhysRevD.86.025022}%
  \BibitemOpen
  \bibfield{author}{%
  \bibinfo {author} {\bibfnamefont{S.}~\bibnamefont{El-Showk}}, \bibinfo
  {author} {\bibfnamefont{M.~F.}\ \bibnamefont{Paulos}}, \bibinfo {author}
  {\bibfnamefont{D.}~\bibnamefont{Poland}}, \bibinfo {author}
  {\bibfnamefont{S.}~\bibnamefont{Rychkov}}, \bibinfo {author}
  {\bibfnamefont{D.}~\bibnamefont{Simmons-Duffin}},\ and\ \bibinfo {author}
  {\bibfnamefont{A.}~\bibnamefont{Vichi}},\ }%
  \bibfield{journal}{%
  \Doi{10.1103/PhysRevD.86.025022}{\bibinfo {journal} {Phys. Rev. D}}\ }%
  \textbf{\bibinfo {volume} {86}},\ \bibinfo {pages} {025022} (\bibinfo {month}
  {Jul}\ \bibinfo {year} {2012})%,\\url{http://link.aps.org/doi/10.1103/PhysRevD.86.025022}%
  \bibAnnoteFile{NoStop}{PhysRevD.86.025022}%
  \bibitem{ElShowk:2012hu}%
  \BibitemOpen
  \bibfield{author}{%
  \bibinfo {author} {\bibfnamefont{S.}~\bibnamefont{El-Showk}}\ and\ \bibinfo
  {author} {\bibfnamefont{M.~F.}\ \bibnamefont{Paulos}}}%
  \  (\bibinfo {year} {2012}),\
  \Eprint{http://arxiv.org/abs/1211.2810}{arXiv:}{1211.2810}%
  \bibAnnoteFile{NoStop}{ElShowk:2012hu}%
%%CITATION = ARXIV:1211.2810;%%
\bibitem{Tomboulis:2009zz}%
  \BibitemOpen
  \bibfield{author}{%
  \bibinfo {author} {\bibfnamefont{E.~T.}\ \bibnamefont{Tomboulis}},\ }%
  \bibfield{journal}{%
  \Doi{10.1142/S0217732309032307}{\bibinfo {journal} {Mod. Phys. Lett. A}}\ }%
  \textbf{\bibinfo {volume} {24}},\ \bibinfo {pages} {2717} (\bibinfo {year}
  {2009})%
  \bibAnnoteFile{NoStop}{Tomboulis:2009zz}%
%%CITATION = MPLAE,A24,2717;%%
\bibitem{DeGrand:2010ba}%
  \BibitemOpen
  \bibfield{author}{%
  \bibinfo {author} {\bibfnamefont{T.}~\bibnamefont{DeGrand}},\ }%
  \bibfield{journal}{%
  \bibinfo {journal} {Phil. Trans. R. Soc. A}\ }%
  \textbf{\bibinfo {volume} {369}},\ \bibinfo {pages} {2701} (\bibinfo {year}
  {2011})%,\ \Eprint{http://arxiv.org/abs/1010.4741}{arXiv:1010.4741 [hep-lat]}%
  \bibAnnoteFile{NoStop}{DeGrand:2010ba}%
%%CITATION = 1010.4741;%%
\end{thebibliography}
\end{document}